\documentclass{emulateapj}

\usepackage{graphicx}
\usepackage{times}

\newcommand{\beq}{\begin{equation}}
\newcommand{\eeq}{\end{equation}}

\def\gs{\mathrel{\lower0.6ex\hbox{$\buildrel {\textstyle >}\over{\scriptstyle \sim}$}}}
\def\ls{\mathrel{\lower0.6ex\hbox{$\buildrel {\textstyle <}\over{\scriptstyle \sim}$}}}

\defcitealias{def+al05}{Paper~I}

\begin{document}

\title{Measuring the Three-Dimensional Structure of Galaxy Clusters. II.
Are clusters of galaxies oblate or prolate?}
\shorttitle{Are galaxy clusters oblate or prolate?}

\author{Mauro Sereno\altaffilmark{1,2,3,4}}
\author{Elisabetta De Filippis\altaffilmark{1,2,5}}
\author{Giuseppe Longo\altaffilmark{1,2,3}}
\author{Mark~W.Bautz\altaffilmark{5}}

\altaffiltext{1}{Dipartimento di Scienze Fisiche, Universit\`{a} degli Studi di Napoli
                `Federico II', Via Cinthia, Compl. Univ. Monte S. Angelo,
                80126 Napoli, Italia}

\altaffiltext{2}{INFN - Sez. Napoli, Compl. Univ. Monte S. Angelo,
                80126 Napoli, Italia; betty@na.infn.it; longo@na.infn.it}

\altaffiltext{3}{INAF - Osservatorio Astronomico di Capodimonte,
                Salita Moiariello, 16, 80131 Napoli, Italia}

\altaffiltext{4}{Institut f\"{u}r Theoretische Physik, Universit\"{a}t Z\"{u}rich,
                Winterthurerstrasse 190, CH-8057 Z\"{u}rich, Schweiz; sereno@physik.unizh.ch}

\altaffiltext{5}{Massachusetts Institute of Technology, Kavli Center for Astrophysics
                and Space Research, Building 37, Cambridge MA 02139 USA;
                mwb@space.mit.edu}

\shortauthors{Sereno et al.}

\date{}

\begin{abstract}
The intrinsic shape of galaxy clusters can be obtained through a
combination of X-ray and Sunyaev-Zeldovich effect observations once
cosmological parameters are assumed to be known. In this paper we
discuss the feasibility of modelling galaxy clusters as either prolate
or oblate ellipsoids. We analyze the intra-cluster medium distribution for a sample of $25$ X-ray selected clusters, with measured Sunyaev-Zeldovich temperature decrements. A mixed population of prolate and oblate ellipsoids of revolution fits
the data well, with prolate shapes preferred on a $\sim 60-76\%$ basis. We observe an excess of clusters nearly aligned along the line of sight, with
respect to what is expected from a randomly oriented cluster
population, which might imply the presence of a selection bias in our
sample. We also find signs that a more general triaxial morphology
might better describe the morphology of galaxy clusters. Additional
constraints from gravitational lensing could disentangle the
degeneracy between an ellipsoidal and a triaxial morphology, and could
also allow an unbiased determination of the Hubble constant.
\end{abstract}


\section{Introduction}

Measuring the three-dimensional shape of any class of astronomical
object is today still an almost unsolved task which has challenged
astronomers for decades. The knowledge of the shape of clusters of
galaxies in particular has fundamental cosmological implications. In a
cold dark matter scenario, smaller clumps aggregate hierarchically
along large-scale perturbations, which are typically highly
anisotropic at their first collapse. Accretion of material might hence
preferentially occur along the filamentary structure within which
larger matter aggregates (i.e. galaxy clusters) are embedded
\citep{wes94}. This could explain the observed tendency of clusters to
be aligned with their nearest neighbor \citep{ca+me80}. Further
processes, such as virialization of the matter aggregates or
self-interaction mechanisms for dark matter, might on the other hand
tend to make such systems more spherical.  Dissipation and gas cooling could result in an average increase of the axis ratios of dark matter halos by $\sim 20-40\%$ in the inner regions, with effects persisting almost to the virial radius \citep{kaz+al04}. The intrinsic shape of structures therefore contains evidence of the formation history of the
large-scale structure and about the nature and mechanisms of interaction of dark matter~\citep[and references therein]{pli+al91,pli+al04}.

The first attempts to determine the three dimensional morphology of
galaxies with only photometric data \citep[and references
therein]{bi+me98} led to the realization that since objects might be
either triaxial or build up a mixture of oblate and prolate shapes, a
purely geometrical and statistical analysis will not produce a unique
solution \citep{bin80}.

First results came from the original work of~\citet{hub26}, who first
determined the relative frequencies with which galaxies of a given
intrinsic ellipticity, oriented at random, will be observed as having
various apparent projected ellipticities. Several studies have
subsequently generalized Hubble's work \citep[and references
therein]{bi+me98}, making the statistical approach of inversion of the
distribution of the apparent shapes of galaxies a widely used method.
This inversion method has several drawbacks, though. First of all, it
is not unambiguous since it requires strong assumptions on the
distribution of the intrinsic shapes of the structures. The
distribution of the apparent flattening is moreover rather
insensitive to modifications in the model assumed for the intrinsic
shapes~\citep{bin80}; different distributions of oblate or prolate
spheroids, or generally triaxial ellipsoids, can consequently be
consistent with the observed distribution of axial ratios. Simplifying
hypotheses, such as requiring structures to be oblate, can hence be
strongly misleading~\citep{be+ca75,ill77}; for a more correct
approach, prolate and triaxial models should also be considered.

While the first investigation by~\citet{hub26} was done on galaxies,
several following studies have faced the inversion method for
different classes of astronomical objects such as poor groups or
clusters of
galaxies~\citep{bi+de81,th+ch01,fa+vi91,noe79,al+ry02,bin80,ryd92,ryd96,pli+al04}.
With the exception of disc galaxies, prolate-like shapes appear to
dominate all cosmic structure on a large scale~\citep{pli+al04}.

In this paper, we are interested in obtaining the three-dimensional
shape of clusters of galaxies. So far, studies on clusters have been
limited to the statistical inversion of the apparent distribution of
axial ratios. \citet{pli+al91} found that the distribution of apparent
axial ratios of 397 Abell clusters, selected from the Lick map of
galaxies, is inconsistent with a population of oblate spheroids.
Similar results are reported by~\citet{bas+al00} and~\citet{det+al95},
who find that a population composed by purely prolate spheroids is
more consistent with, respectively, a sample of 903 clusters from the
APM catalog and with a set of 99 low redshift Abell clusters, better
than a purely oblate population. While the above studies were
performed uniquely on optical data, \citet{coo00} used the X-ray
isophotal axial ratios of a sample of 58 clusters by~\citet{moh+al95};
a population of purely oblate ellipsoids is again ruled out. A quite different approach was followed in \citet{paz+al06}, where observational data from 2PIGG and SDSS group catalogues were compared to projected shapes of numerically simulated haloes. Haloes turned out to be preferentially prolate, with more massive groups tending to show more elongated shapes.

The above listed results are based on the controversial assumption
that objects are randomly oriented in space. Furthermore, tests for
possible intrinsic triaxial or mixed populations have to date been far
from conclusive. Several works in the past few years have discussed
how multi-wavelength observations of clusters of galaxies can be used
to uncover the 3-D structure of clusters of
galaxies~\citep{zar+al98,zar+al01,reb00,fo+pe02}. In a companion
paper~\citep{def+al05} (hereafter,~\citetalias{def+al05}), we have
considered the theoretical capability of combined X-ray,
Sunyaev-Zeldovich effect (SZE) and gravitational lensing observations
to determine the 3-D morphology of relaxed galaxy clusters and,
jointly, to constrain cosmological parameters.  Due to a lack of
gravitational lensing data, we have then used X-ray and SZE
measurements of a sample of 25 clusters to constrain their triaxial
structure by assuming the concordance cosmological $\Lambda$CDM model,
which is determined to an unprecedented accuracy thanks to an
impressive body of evidence coming from several cosmological
tests~\citep{wan+al00,teg+al04}.

In this paper, we test the accuracy to which ellipsoids of revolution
can describe the sample of clusters of galaxies we studied
in~\citetalias{def+al05}. Following a similar empirical approach, we
use combined X-ray and SZE observations to directly probe the 3-D
structure of clusters in the sample, without any assumptions on their
statistical properties. The paper is organized as follows. In
\S~\ref{sec:elli} we discuss how the shape of an ellipsoid of
revolution can be obtained from combined X-ray and SZE measurements.
In \S~\ref{sec:intr_distr} we introduce our cluster sample and derive
the intrinsic distributions of the axial ratios and inclination angles
for the cluster sample, together with a comparison between these results and those obtained in ~\citetalias{def+al05} assuming a triaxial morphology aligned along the line of sight. In \S~\ref{sec:inversion} we perform a statistical inversion of the
observed distribution of projected axial ratios. In
\S~\ref{sec:syst} some systematics are considered. Further information
that can be obtained with additional constraints from gravitational
lensing observations is discussed in
\S~\ref{sec:lens}. \S~\ref{sec:conc} is devoted to
conclusions and final considerations. In Appendix~\ref{sec:appendiceA}, we
provide some mathematical details on the two-dimensional projection of
an ellipsoid.

\section{Ellipsoids of revolution}
\label{sec:elli}

In this paper we assume that galaxy clusters have an axisymmetric gas
density distribution. The intrinsic distribution of the intra-cluster
medium (ICM), assumed to be constant on similar, concentric spheroidal
ellipsoids, is characterized by the ratio of the major to minor axes
$e_\mathrm{int}(>1)$, and by the inclination angle $i$, between the
line of sight (LOS) of the observer and the polar axis. In an
intrinsic orthogonal coordinate system centred on the cluster's
barycentre and whose coordinates are aligned with its principal axes,
a spheroidal ICM profile can be described by only one radial variable
$\zeta$,
\beq
\zeta^2 \equiv \sum_{i=1}^3 e_i^2 x_{i,\mathrm{int}}^2 .
\label{eq:rad_var}
\eeq
If we take the axis of symmetry aligned with the third coordinate,
$e_i= \left\{ e_\mathrm{int}, e_\mathrm{int},1 \right\}$ for a prolate
model and $e_i=\left\{ 1,1,e_\mathrm{int} \right\}$ for an oblate
model.

For an intrinsic 3-D spheroidal morphology, the two-dimensional
observed quantities, projected on the plane of the sky (POS), are
constant on concentric ellipses. The axial ratio of the major to minor
axes of the observed projected isophotes, $e_\mathrm{proj}$, can be
expressed as a function of the cluster intrinsic
parameters~\citep{fab+al84},
\beq
\label{obl2}
e_{\rm proj} = \left\{
\begin{tabular}{ll}
$\sqrt{\cos^2 i+ e_\mathrm{int}^2 \sin^2 i}$            & {\rm prolate
case}
\\ $ {\displaystyle  \frac{e_\mathrm{int}}{\sqrt{e_\mathrm{int}^2\cos^2 i + \sin^2 i}} }$ & {\rm oblate case}  \\
\end{tabular}
\right.
\eeq

In~\citetalias{def+al05}, we discussed how the elongation along the
LOS of a cluster of galaxies, $e_\mathrm{ LOS}$, can be determined by
combining X-ray and SZE observations if a reliable estimate of the
Hubble constant is provided. For a spheroid inclined with respect to
the LOS, as we assume clusters to be shaped in this paper, $e_\mathrm{
LOS}$ represents the ratio of the size of the spheroid along the LOS,
measured through its barycentre, to the size of the major axis
projected in the POS. A value of $e_\mathrm{LOS}$ greater than one
corresponds to a cluster elongated along the LOS, and vice versa for $e_\mathrm{LOS}<1$. For
the cluster plasma density we assume an isothermal
$\beta$-distribution:
\begin{equation}
\label{eq:tri0}
n_\mathrm{e} (\zeta)  \propto \left( 1+ \frac{\zeta^2}{r_{\rm c}^2}
\right)^{-3\beta/2},
\end{equation}
where $r_{\rm c}$ is the core radius and $\beta$ the slope parameter.
For such a distribution, $e_\mathrm{LOS}$ can then be written as
~\citepalias{def+al05}:
\begin{eqnarray}
\label{eq:obl7}
e_\mathrm{LOS} & \equiv & \left[ \frac{\Delta T_0^2}{S_\mathrm{X0}}
\left( \frac{m_\mathrm{e} c^2}{k_\mathrm{B} T_\mathrm{e} } \right)^2
\frac{\Lambda_\mathrm{e}\mu_\mathrm{e}/\mu_\mathrm{H} }{4 \pi^{3/2}f(\nu,T_{\rm e})^2
T^2_\mathrm{CMB} \sigma_{\rm T}^2 (1+z_{\rm c})^4}  \right. \nonumber
\\ & {\times} &
\left. \left( \frac{\Gamma [3 \beta /2 ]}{\Gamma [3 \beta /2 -1/2 ]}  \right)^2
\frac{\Gamma [3 \beta -1/2 ]}{\Gamma [3 \beta ]} \frac{1}{ \theta_\mathrm{c,proj} } \right]
\frac{1}{ D_\mathrm{c}} .
\end{eqnarray}
where $z$ is the cluster redshift, $D_\mathrm{c}$ the angular diameter
distance to the cluster, $\Delta T_0$ the central SZE temperature
decrement, $\theta_\mathrm{c,proj}$ the projected angular major core
radius, $T_{\rm e}$ the temperature of the ICM, $k_{\rm B}$ the
Boltzmann constant, $T_{\rm CMB} = 2.728^{\circ}$K \citep{fix+al96} the
temperature of the cosmic microwave background radiation, $\sigma_{\rm
T}$ the Thompson cross section, $m_{\rm e}$ the electron mass, $\mu
m_{\rm p}$ the mean particle mass of the gas, $c$ the speed of light
in vacuum, $\Lambda_{e}$ the X-ray cooling function of the ICM in the
cluster rest frame, $S_\mathrm{X0}$ the central X-ray surface
brightness and $f(\nu, T_{\rm e})$ accounts for frequency shift and
relativistic corrections. In a flat
Friedmann-Lema\^{\i}tre-Robertson-Walker model of universe filled with
dust, $\Omega_\mathrm{M0}$, and with a non null cosmological constant,
the angular diameter distance to a source at a redshift $z$ is
\citep{ser+al01}
\beq
\label{eq:crit3}
D_\mathrm{c} = \frac{c}{H_0} \frac{1}{1+z} \int_0^z
\frac{dx}{ \sqrt{ \Omega_\mathrm{M0} (1+x)^3+1-\Omega_\mathrm{M0} } }.
\eeq
We assume a flat $\Lambda$CDM model with a Hubble constant
$H_0=70^{+4}_{-3}$~km~s$^{-1}$Mpc$^{-1}$ and $\Omega_{\rm M0}= 0.30 {\pm}
0.04$~\citep{teg+al04}. As well known, if the Hubble constant is not assumed from independent observational constraints, then the length of a cluster along the LOS can be known from SZE and X-ray observations only up to a multiplicative factor \citep{coo98}. When we use experimental data provided with
asymmetric uncertainties, corrections as given by~\cite{dag04} are
applied to obtain unbiased estimates of mean and standard deviation.

In terms of $e_\mathrm{proj}$ we can also
write~\citep{def+al05,don+al03}
\beq
\label{obl3}
e_\mathrm{LOS} =
\left\{
\begin{tabular}{ll}
$\displaystyle{ \frac{ \sqrt{e_{\rm proj}^2- \cos^2 i}} {e_{\rm proj}^2 \sin i}} $ &
{\rm prolate case} \\ $\displaystyle{ \frac{\sqrt{1- e_{\rm proj}^2 \cos^2 i}}{\sin
i} }$ & {\rm oblate case}
\end{tabular}
\right.
\eeq
For clusters aligned along the LOS (i.e. $i = 0$) both prolate and
oblate distributions exhibit circular isophotes and $e_{\rm LOS}$ can
be expressed as a function of the cluster intrinsic ellipticity as
follows:
\beq
\label{simp4}
e_\mathrm{LOS} =
\left\{
\begin{tabular}{ll}
$e_\mathrm{int}$          & {\rm prolate case}   \\ $\displaystyle{ \frac{1}{e_{\rm
int}}  }$ & {\rm oblate case}.  \\
\end{tabular}
\right.
\eeq

If we assume the cluster to be either prolate or oblate
Eqs.~(\ref{obl2},~\ref{obl3}), enable us to determine the intrinsic
geometry of the cluster. In the prolate case, we obtain
\begin{eqnarray}
e_\mathrm{int}& =& e_\mathrm{LOS} e_\mathrm{proj}^2;
\label{sys_pro_1}
\\ i
&= & {\pm}
\arccos \left[
\sqrt{\frac{e_\mathrm{int}^2 - e_\mathrm{proj}^2}{e_\mathrm{int}^2 - 1}}
\right], \label{sys_pro_2}
\end{eqnarray}
while in the oblate case:
\begin{eqnarray}
e_\mathrm{int}& =& \frac{e_\mathrm{proj}}{e_\mathrm{LOS}}
\label{sys_obl_1} \\ i & = & {\pm} \arccos
\left[
\sqrt{\frac{e_\mathrm{LOS}^2 - 1}{e_\mathrm{LOS}^2- e_\mathrm{proj}^2}}
\right]. \label{sys_obl_2}
\end{eqnarray}
We are dealing with observed projected quantities which are invariant
under a reflection through the POS. Hence we are not able to determine
which extremity of the cluster is pointing towards the observer.

\section{Intrinsic distributions through combined X-ray and SZE measurements}
\label{sec:intr_distr}

\tabletypesize{\scriptsize}
\def\arraystretch{1.0}
\begin{deluxetable*}{llllll}
\tablecolumns{6}
\tablewidth{0pt}
\tablecaption{Intrinsic Parameters of Sample Clusters}
\tablehead{
\colhead{}       &\colhead{}    &\colhead{}&\colhead{$q_\mathrm{int}$}&
\colhead{$i$}&\colhead{$q_\mathrm{int}$}  \\
\colhead{Cluster}&\colhead{$z$}&\colhead{Shape}
&\colhead{(ellips.)}&\colhead{(deg)}&\colhead{(triax.)}  }
\startdata
MS 1137.5+6625  &       0.784   &       pro     &       0.49    $\pm$
0.21            &       16      $\pm$   9       & $0.56\pm{0.21}$\\ MS
0451.6-0305  &       0.550   &       obl &       0.71    $\pm$   0.16
&       58      $\pm$   43      & $0.76\pm{0.06}$ \\
--              &       --      &       pro     &       0.63    $\pm$   0.14    &       56      $\pm$   13      &       --\\
Cl 0016+1609~\tablenotemark{a}    &       0.546   &       pro     &
0.55    $\pm$ 0.13            &       26      $\pm$   10      &
$0.67\pm{0.14}$\\ RXJ1347.5-1145~\tablenotemark{a}  &       0.451   &
obl & 0.68 $\pm$   0.16    &       84      $\pm$   117     &
$0.69\pm{0.09}$\\
--              &       --      &       pro     &       0.48    $\pm$   0.11    &       35      $\pm$   12      &       --\\
A 370           &       0.374   &       pro     &       0.35    $\pm$
0.13            &       27      $\pm$   12      & $0.55\pm{0.15}$ \\
MS 1358.4+6245~\tablenotemark{a}  &       0.327   &       obl
& 0.55 $\pm$   0.14    &       52      $\pm$   8       &
$0.71\pm{0.06}$
\\
A 1995          &       0.322   &       obl     &
0.75    $\pm$   0.18 & 64      $\pm$   35      & $0.81\pm{0.09}$ \\
--              &       --      &       pro     &       0.70    $\pm$   0.16    &       46      $\pm$   27      &       --\\
A 611           &       0.288   &       pro     &       0.73    $\pm$
0.26            &       35      $\pm$   32      & $0.84\pm{0.19}$ \\ A
697           &       0.282   &       pro                     & 0.44
$\pm$   0.12    &       26      $\pm$   10      & $0.60\pm{0.14}$ \\ A
1835~\tablenotemark{a}          &       0.252   &       pro     & 0.56
$\pm$   0.08 & 29      $\pm$   7       & $0.70\pm{0.09}$ \\ A
2261~\tablenotemark{a} & 0.224 &       pro &       0.63    $\pm$ 0.16
&       10 $\pm$ 6 & $0.65\pm{0.16}$ \\ A 773           & 0.216   &
pro     & 0.32 $\pm$   0.09            &       14      $\pm$ 5 &
$0.40\pm{0.11}$ \\ A 2163~\tablenotemark{a}          &       0.202   &
pro & 0.62    $\pm$   0.13 &       28      $\pm$   10      &
$0.73\pm{0.12}$
\\ A 520           & 0.202   &       obl & 0.54    $\pm$   0.20    &
23      $\pm$ 10      & $0.56\pm{0.08}$ \\ A 1689~\tablenotemark{a}
& 0.183   & obl & 0.85    $\pm$   0.12    &       65      $\pm$   44 &
$0.88\pm{0.06}$ \\
--              &       --      &       pro     &       0.79    $\pm$   0.11    &       46      $\pm$   23      &       --\\
A 665           &       0.182   &       obl     &       0.58    $\pm$
0.25            &       47      $\pm$   13      & $0.71\pm{0.15}$\\ A
2218          &       0.171   &       pro                     & 0.54
$\pm$   0.18    &       22      $\pm$   11      & $0.64\pm{0.18}$ \\ A
1413~\tablenotemark{a}          &       0.142   &       obl     & 0.64
$\pm$   0.17 & 73      $\pm$   33      & $0.68\pm{0.09}$ \\
--              &       --      &       pro     &       0.49    $\pm$   0.13    &       37      $\pm$   15      &       --\\
A 2142~\tablenotemark{a}          &       0.091   &       obl     &
0.63    $\pm$ 0.13            &       79      $\pm$   42      &
$0.65\pm{0.07}$\\
--              &       --      &       pro     &       0.43    $\pm$   0.09    &       34      $\pm$   10      &       --\\
A 478~\tablenotemark{a}           &       0.088   &       pro     &
0.34    $\pm$ 0.14            &       23      $\pm$   12      &
$0.51\pm{0.17}$ \\ A 1651~\tablenotemark{a}          &       0.084   &
pro & 0.31 $\pm$   0.16    &       12      $\pm$   7       &
$0.37\pm{0.19}$\\ A 401           &       0.074   &       pro     &
0.46    $\pm$   0.08 & 25      $\pm$   6       & $0.60\pm{0.10}$\\ A
399           & 0.072 &       obl &       0.50    $\pm$   0.14    & 38
$\pm$   5 & $0.58\pm{0.04}$ \\ A 2256          &       0.058   & pro
& 0.54 $\pm$   0.14            &       34      $\pm$   14 &
$0.72\pm{0.11}$
\\ A 1656          &       0.023   &       pro & 0.70    $\pm$   0.28
&       33      $\pm$   30      & $0.81\pm{0.21}$ \\
\enddata{}
\tablenotetext{a}{Cooling flow clusters.}
\tablecomments{For ambiguous clusters, both prolate and oblate solutions are listed.
$z$ is the redshift, $q_\mathrm{int}$ (ellips.) the cluster maximum
axis ratio under an ellipsoidal assumption, $i$ the inclination angle,
$q_\mathrm{int}$ (triax.) the cluster maximum axis ratio under a
triaxial assumption.}
\label{tab:sample}
\end{deluxetable*}

In this section, we apply the formalism described in \S~\ref{sec:elli}
to compute the three-dimensional intrinsic elongation and inclination
for the same cluster sample used in \citetalias{def+al05}, for which a
combined X-ray and SZE data set was available. The sample consists of
25 clusters of galaxies, 18 from~\citet{ree+al02} and 7
from~\citet{mas+al01}. Details on the sample and on the selection
criteria can be found in~\citet{mas+al01,ree+al02,def+al05}. Basic
data for our $25$ clusters, including previously published redshift,
plasma temperature and SZE decrement information are presented in
Table~1 of~\citetalias{def+al05}. In~\citetalias{def+al05}, we also
measured projected axial ratios, $e_\mathrm{proj}$, for all clusters
in the sample through an X-ray morphological analysis based on high
resolution {\it Chandra} and {\it XMM-Newton} observations; measured
values are listed in Table~2 of \citetalias{def+al05}.

Through a combined analysis of both X-ray and SZE observations, we can
infer new information about intrinsic elongation and inclination of
the clusters in the sample. We then model these distributions with a
kernel empirical estimator~\citep{vio+al94,ryd96}. Given a sample of
$N$ measured quantities $\eta_i$, the kernel estimator of the
distribution is a continuous function given by:
\beq
\label{kern1}
N_\eta (\eta)=\frac{1}{N h} \sum_{i=1}^N K\left(
\frac{\eta - \eta_i }{h}\right)
\eeq
where $K$ is the kernel function. When densities take values in the
range $0 \leq \eta \leq \eta_\mathrm{max} $, the kernel can be written
as:
\begin{eqnarray}
\label{kern2}
K_{\rm ref}(\eta, \eta_i, h) & = & K_{\rm Gau}\left(
\frac{\eta - \eta_i}{h}\right)+ K_{\rm Gau}\left(
\frac{\eta + \eta_i}{h}\right)\\
& + & K_{\rm Gau}\left( \frac{2\eta_\mathrm{max}- \eta -
\eta_i}{h}\right) \nonumber
\end{eqnarray}
where $K_{\rm Gau}$ is a Gaussian kernel given by:
\beq
K_{\rm Gau} (x)=\frac{1}{\sqrt{2 \pi}} e^{-x^2/2}.
\eeq
When using the kernel in Eq.~(\ref{kern2}), Gaussian tails extending
to negative values or to values greater than $\eta_\mathrm{max}$, are
folded back between 0 and $\eta_\mathrm{max}$. The kernel width $h$
can be approximated by $h \simeq 0.9 A N^{-0.2}$, where $A$ is the
smallest quantity between the standard deviation of the sample and the
interquartile range divided by 1.34 \citep{vio+al94}.

In order to account for the effect of the finite sample size, we
derive the confidence levels from a bootstrap re-sampling of the data
set. From the distribution obtained with the original $N$ data points,
we draw with replacement a new set of $N$ data points which are then
used to create a new bootstrap estimate of the kernel estimator,
$N_\eta$. Confidence intervals can then be assigned to each value of
$\eta$, by finding the values of $N_\eta$ that lie above (below) the
required upper (lower) confidence limit. Errors in the measured value
of $\eta$ induce a further uncertainty in the estimated distribution.
This effect can be modelled with a kernel width given by:
\beq
h^{'}_i=\sqrt{h^2+\Delta \eta_i^2}
\label{eq:h_primo}
\eeq
which changes for each data point \citep{ryd96}.

\begin{figure}
        \resizebox{\hsize}{!}{\includegraphics{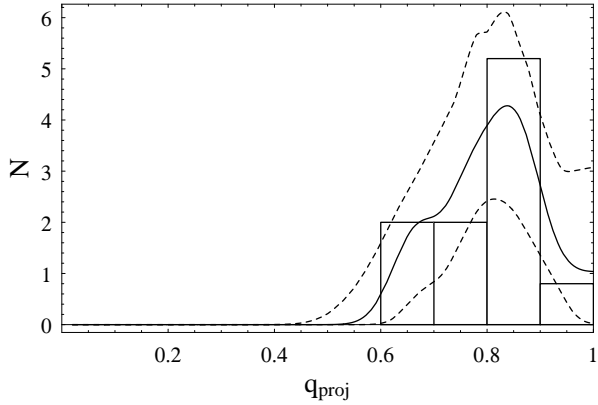}}
        \caption{Normalized distribution of apparent axial ratios. Solid and dashed lines show the
        non-parametric kernel estimate and the 2-$\sigma$ confidence range using a
        bootstrap procedure.}
        \label{fqHB_istog_2Sigma}
\end{figure}

In what follows, we use the inverse of the intrinsic and projected
axial ratios: $q_\mathrm{int} \equiv 1/e_\mathrm{int}$ and
$q_\mathrm{proj} \equiv 1/e_\mathrm{proj}$. Such a formalism allows
axial ratios to be limited to the finite range $\left(0, 1\right]$.
The histogram in Fig.~\ref{fqHB_istog_2Sigma} represents the observed
distribution of the projected axial ratios. Confidence limits are
built up with $10^3$ bootstrap re-samplings. Estimates of the observed
projected ellipticity from both {\it Chandra} and {\it XMM-Newton}
satellites are so accurate that for the distribution of observed axial
ratios $h^{'}_i \simeq h$. The
measured mean ellipticity is $q_\mathrm{proj} =0.80{\pm}0.02$; the value
of the median is instead 0.81. Reported uncertainties represent the
standard deviation of the mean. These values are in good agreement with the average projected axial ratio expected if dark matter halos of clusters do have mass distributions as revealed by numerical simulations \citep{wa+fa04}. The solid and the dashed lines in
Fig.~\ref{fqHB_istog_2Sigma} show the non-parametric kernel estimate
and the 2-$\sigma$ confidence range, respectively, using the bootstrap
procedure.

\subsection{Axial ratios}

\begin{figure}
        \resizebox{\hsize}{!}{\includegraphics{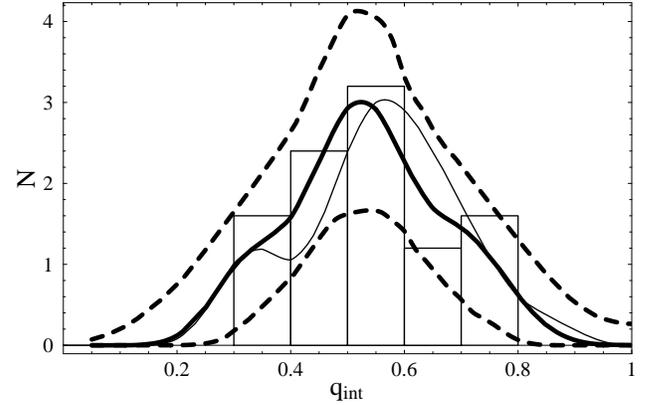}}
        \caption{Distribution of intrinsic axial ratios. Histogram and thick lines
        are computed assuming a prolate solution for all ambiguous clusters.
        Thick-solid and thick-dashed lines show
        the non-parametric kernel estimate and the 2-$\sigma$ confidence range using a
        bootstrap procedure. The thin line is the kernel parametric estimate assuming the oblate
        solution for all ambiguous clusters. Histogram and all distributions are normalized to unity.}
        \label{fqInt_istog_2Sigma}
\end{figure}

In~\citetalias{def+al05}, the approach used to calculate the
elongation along the LOS is discussed in detail; estimated values of
$e_\mathrm{LOS}$ are listed there in Table 4. Once $e_\mathrm{LOS}$
and $e_\mathrm{proj}$ are known, both systems
Eqs.~(\ref{sys_pro_1},~\ref{sys_pro_2}) and
Eqs.~(\ref{sys_obl_1},~\ref{sys_obl_2}) for prolate and oblate
ellipsoids, respectively, can be solved. Given each set of
observational constraints, a cluster can admit either only one
solution (prolate or oblate) or both. By solving the equation systems,
we find that of the 25 clusters in our sample, 15 are compatible only
with a prolate morphology, 4 only with an oblate one, while the
remaining 6 clusters (hereafter the ``ambiguous clusters'') are
compatible with both the prolate and the oblate assumption. The
inferred intrinsic parameters are listed in Table~\ref{tab:sample}.
For comparison, in the last column of Table~\ref{tab:sample} we list
the values of $q_{\rm int}$ as computed in~\citetalias{def+al05}
assuming clusters as triaxial structures with null inclination.

To analyze the inferred distribution and to model it with the kernel
estimator, we first assume that all of the ambiguous clusters are
prolate (Case I) and then that they are all oblate (Case II). These
are both extreme cases, but the intrinsic distribution turns out not
to be very sensitive to this assumption. In
Fig.~\ref{fqInt_istog_2Sigma} we plot the distribution of the
intrinsic axial ratios. Also in this case, confidence ranges have been
determined using the bootstrap procedure with $10^3$ re-samplings. It
can be seen that the kernel estimate does not change significantly if
all ambiguous clusters are assumed to be all prolate or oblate (solid
thick and thin line, respectively). The distribution obtained in Case
II (thin line) is well within the 2-$\sigma$ confidence level of the
one obtained in Case I (thick-dashed lines). In Case I (II), the mean
intrinsic axial ratio is $0.54 {\pm} 0.03$ ($0.56 {\pm} 0.03$), while the
median is 0.54 (0.56).

\subsection{Inclination angles}

\begin{figure}
        \resizebox{\hsize}{!}{\includegraphics{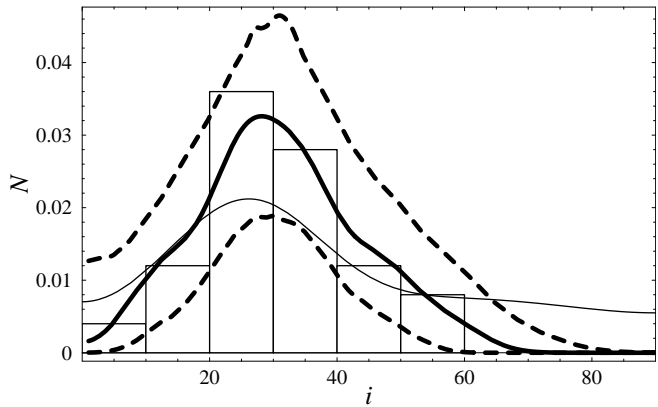}}
        \caption{Distribution of inclination angles. Histogram and lines are
        the same as in Fig.~\ref{fqInt_istog_2Sigma}}
        \label{fTheta_istog_2Sigma}
\end{figure}

\begin{figure}
        \resizebox{\hsize}{!}{\includegraphics{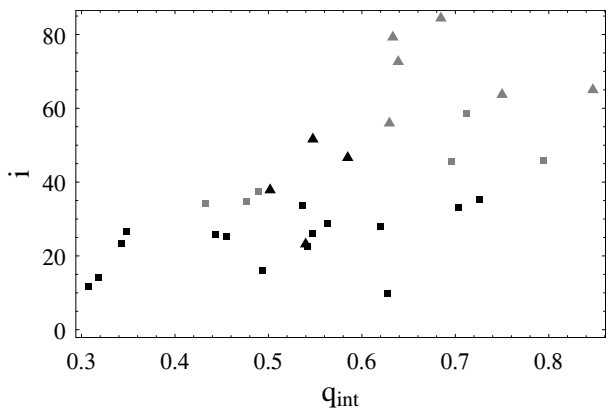}}
        \caption{Orientation angle versus the intrinsic ellipticity for all clusters in
        our sample. Boxes
        and triangles represent prolate and oblate clusters, respectively. Both solutions
        for ambiguous clusters are plotted in gray.
        }
        \label{qInt_vs_Inc_all_cases}
\end{figure}

%\begin{figure}
%        \resizebox{\hsize}{!}{\includegraphics{qInt_vs_Inc_caseI.eps}}
%        \caption{Intrinsic ellipticity versus orientation angle for our sample in case I.
%        }
%        \label{qInt_vs_Inc_caseI}
%\end{figure}

The polar axis of clusters is usually assumed to be randomly oriented
with respect to the line of sight. Under this assumption a fraction
$\sin i\ di$ of all clusters would have their symmetry axes directed
within $di$ of angle $i$ with respect to the line of
sight~\citep{bi+me98}. The mean value of such a distribution would be
$\langle i
\rangle_\mathrm{random} \sim 57^{^{\circ}}$.
We instead observe a slightly different picture. The measured
distribution of the inclination angles is shown in
Fig.~\ref{fTheta_istog_2Sigma}. The mean inclination angle is $ 32 {\pm}
3\deg$ ($37 {\pm} 4\deg$) for Case I (II), whereas the median is $29\deg$
for both cases. Observed clusters show an excess of clusters with
their polar axes closely aligned with the line of sight, and a strong
deficit of clusters with high values of $i$. The distribution plotted
in Fig.~\ref{fTheta_istog_2Sigma} has been obtained without
considering the error $\Delta i$ when calculating the kernel widths.
Using $h_i^{'}$ instead of $h_i$ causes a further broadening of the
distribution; both the excess of small values of $i$ and the
discrepancy with the assumption of a random distribution still hold.

Fig.~\ref{qInt_vs_Inc_all_cases} shows a plot of the orientation angle
versus the intrinsic ellipticity for all clusters in our sample. The
most elongated clusters (small values of $q_{\rm int}$) turn out to be
prolate nearly aligned with the line of sight. This is most probably
due to a bias in the selection of the clusters in the sample. Clusters
with low X-ray luminosity, which are stretched along the line of sight
are preferentially included in a surface brightness limited survey.
Both the ellipticity and the inclination inferred for ambiguous
clusters are quite sensitive to the assumed shape.  The oblate solution
implies a more spherical geometry (larger values of $q_\mathrm{int}$)
and a more inclined polar axis than does the prolate solution. In
fact, ambiguous clusters can have inclination angles $\gs 60\deg$ only
if they are oblate.

\subsection{Rotational versus triaxial spheroids}
\label{sec:rto_vs_ell}

\begin{figure}
        \resizebox{\hsize}{!}{\includegraphics{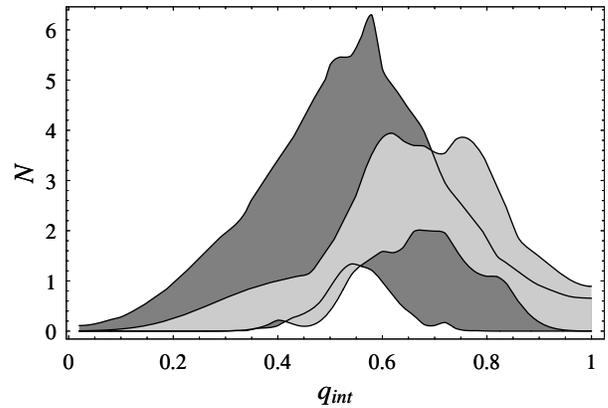}}
        \caption{3-$\sigma$ confidence bands of the intrinsic axis ratio distribution,
        as determined through combined X-ray and SZE measurements with the assumption of an inclined ellipsoidal morphology (dark shaded
        region, Case I), and as determined under the assumption of
        triaxial morphology aligned along the LOS (light shaded region).
        Confidence bands are found by bootstrap re-sampling.}
        \label{fqInt_vs_fqIntTri_3Sigma}
\end{figure}

While in~\citetalias{def+al05} we assumed galaxy clusters to be
triaxial ellipsoids aligned along the LOS, in this paper we have
relaxed the assumption of the null inclination, allowing clusters to
be inclined along the LOS, but we have put a slightly tighter
constraint on their morphology, forcing them to be ellipsoids of
revolution.

We observe a tight concordance between the results obtained in the two
analyses. Estimated values of $q_{\rm int}$ computed under the two
assumptions are in agreement in most cases (see
Table~\ref{tab:sample}). The case of a cluster compatible only with
the prolate (oblate) shape corresponds to a triaxial cluster with the
major (minor) axis oriented along the line of sight. Ambiguous
clusters correspond to triaxial shapes having the middle axis directed
along the line of sight. The two distributions of the intrinsic axis
ratio as obtained either in the present analysis or in
\citetalias{def+al05} are represented in
Fig.~\ref{fqInt_vs_fqIntTri_3Sigma} (dark and light shaded regions,
respectively). As a general trend, we observe that galaxy clusters
turn out to be slightly more elongated when modelled as ellipsoids of
revolution than in the case of triaxial ellipsoids (showing an average
ratio of the minor to the major axis of $\sim 0.63 {\pm} 0.03$). This
effect is due to the capability of triaxial shapes to fit observed
data without the need of extreme axis ratios.

\section{Deprojected distribution}
\label{sec:inversion}

\begin{figure}
        \resizebox{\hsize}{!}{\includegraphics{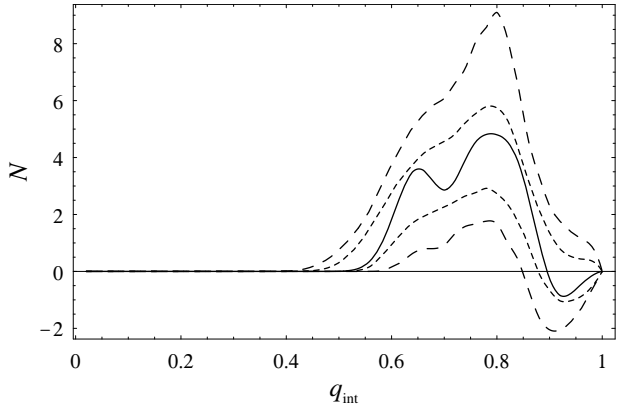}}
        \caption{Deprojected distribution of intrinsic axis ratios assuming that
        all clusters are randomly oriented prolate spheroids (solid line).
        Short-dashed
        and long-dashed lines show the 1 and 2-$\sigma$ confidence limits,
        respectively, estimated using a bootstrap re-sampling technique.}
        \label{NPro_1and2Sigma}
\end{figure}

\begin{figure}
        \resizebox{\hsize}{!}{\includegraphics{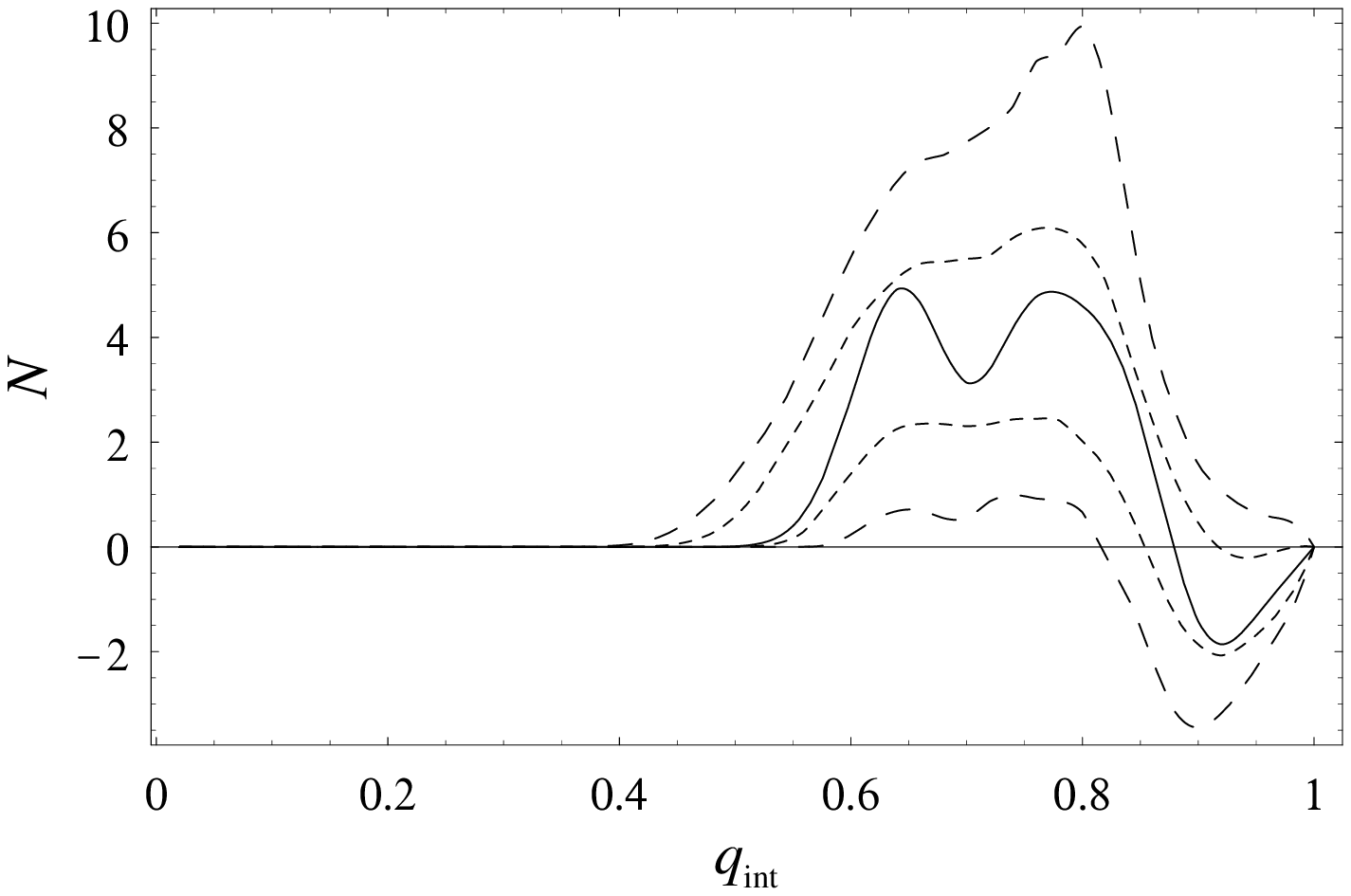}}
        \caption{Deprojected distribution of intrinsic axial ratios assuming that
        all clusters are randomly oriented oblate spheroids (solid line).
        Short-dashed
        and long-dashed lines show the 1 and 2-$\sigma$ confidence limits,
        respectively, estimated using a bootstrap re-sampling technique.}
        \label{NObl_1and2Sigma}
\end{figure}

\begin{figure}
        \resizebox{\hsize}{!}{\includegraphics{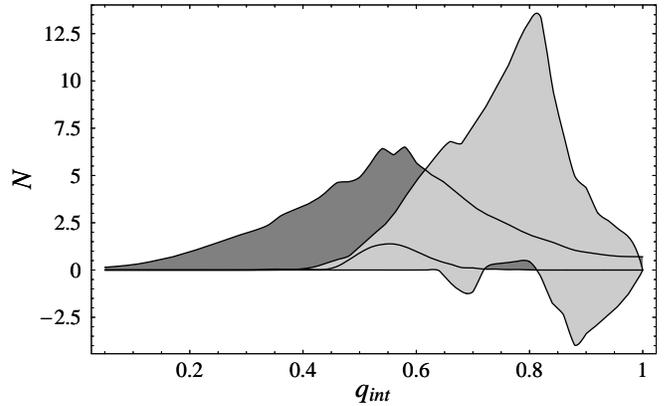}}
        \caption{3-$\sigma$ confidence bands of the intrinsic axial ratios, as
        determined from combined X-ray and SZE measurements (dark shaded region),
        and of the deprojected distribution of observed axial ratios, assuming
        that clusters
        are randomly oriented prolate spheroids (light shaded region). Confidence
        bands are found by bootstrap re-sampling.}
        \label{fqInt_vs_NPro_3Sigma}
\end{figure}

In this section, we compute the distribution of the intrinsic axis
ratio with a third method. The observed projected axial ratio
distribution, $\hat{f} (q_\mathrm{proj})$, can in fact be inverted
under appropriate assumptions in order to obtain the intrinsic
distribution of the axis ratios. If clusters were all randomly
oriented prolate ellipsoids, the intrinsic distribution
$\hat{N}_\mathrm{Pro}$ could be written as \citep{ryd96}
\beq
\hat{N}_\mathrm{Pro} (q_\mathrm{int}) = \frac{2
\sqrt{ 1-q_\mathrm{int}^2} }{\pi q_\mathrm{int}}
\int_0^{ q_\mathrm{int} } \frac{d}{d q} \left(q^2 \hat{f}(q) \right)
\frac{d q}{ \sqrt{q_\mathrm{int}^2-q^2} }.
\eeq
If, on the contrary, all systems are assumed to be randomly oriented
oblate ellipsoids, then the intrinsic distribution can be written as:
\beq
\hat{N}_\mathrm{Obl} (q_\mathrm{int}) =
\frac{ 2 q_\mathrm{int} \sqrt{ 1-q_\mathrm{int}^2} }{\pi}
\int_0^{ q_\mathrm{int} } \frac{d}{d q}
\left( \frac{ \hat{f}(q) }{q} \right)
\frac{d q}{ \sqrt{q_\mathrm{int}^2-q^2} } .
\eeq
To take into account the effect of the finite sample size, confidence
levels for $\hat{N}_\mathrm{Pro}$ and $\hat{N}_\mathrm{Obl}$ have been
derived using a bootstrap re-sampling procedure, see
\S~\ref{sec:intr_distr}. The bootstrap estimates of $\hat{f}$ are then
inverted to compute estimates of the intrinsic distributions
$\hat{N}_\mathrm{Pro}$ and $\hat{N}_\mathrm{Obl}$. Confidence
intervals are assigned to each value of $q_\mathrm{int}$ by finding
values of $\hat{N}_\mathrm{Pro}$ and $\hat{N}_\mathrm{Obl}$ that lie
above (below) some upper (lower) confidence limits. Accurate
confidence limits were built using $10^3$ bootstrap re-samplings. The
resulting intrinsic axial ratio distributions are plotted in
Figs.~\ref{NPro_1and2Sigma} and \ref{NObl_1and2Sigma}; these were
obtained assuming purely prolate and oblate morphologies,
respectively. Negative values in the inverted distributions are
evidence against the underlying hypotheses~\citep{ryd96,coo00}. The
assumption that all clusters are randomly oriented prolate ellipsoids
turns out to be consistent with the observed distribution of projected
axial ratios, see Fig.~\ref{NPro_1and2Sigma}. On the other hand the
hypothesis that all clusters are purely oblate is discarded at the
1-$\sigma$ confidence limit since the resulting intrinsic distribution
is negative for high values of $q_\mathrm{int}$, see
Fig.~\ref{NObl_1and2Sigma}. However, at the 2-$\sigma$ confidence
level, such an hypothesis cannot be rejected, given the observed
ellipticities of our sample clusters.

We now compare the intrinsic distributions of axis ratios obtained
using our deprojection technique through X-ray and SZE observations,
with the above inverted distribution. We have taken into account the
3-$\sigma$ confidence levels of the intrinsic axis ratio distribution
obtained with combined X-ray/SZE method, estimated independently for
Case I and II  in \S~\ref{sec:intr_distr}. We have then obtained the
concordance region by holding all points compatible with at least one
of the cases. The resulting region is plotted in
Fig.~\ref{fqInt_vs_NPro_3Sigma} in dark gray. The light gray region
shows the inverted distribution of observed axial ratios, assuming
that clusters are randomly oriented prolate spheroids. With respect to
our X-ray/SZE technique, the deprojection method leads to a deficit of
nearly spherical structures (low values of $q_{\rm int}$) and to an
overabundance of highly elongated clusters. The same discrepancy is
observed if the distribution of observed axial ratios is inverted
assuming that clusters are randomly oriented oblate spheroids. Both
assumptions that clusters are either all randomly oriented prolate or
oblate spheroids do not provide suitable explanation for the intrinsic
distribution of our sample.

\section{Gravitational lensing constraints}
\label{sec:lens}

As already discussed in \citetalias{def+al05}, a third independent observational input can both break the degeneracy between oblate and prolate models and determine cosmological parameters. Such an input can be provided by observations of strong lensing events.  Whereas the ICM distribution traces the gravitational potential, gravitational lensing directly maps the cluster total mass. We remark that since we are dealing with an ellipsoidal ICM distribution, the isodensity contours will not in general be ellipsoidal, this being approximately true only for a small value of the ellipticity \citep{fo+pe02}. Observations of strong lensing events can put accurate constraints on the convergence, i.e. the dimensionless projected surface density of the total mass distribution of the lens. Assuming hydrostatic equilibrium and a $\beta$-model for the ICM, the central value of the convergence reads \citep[Eq.~(C7)]{def+al05} :
\beq
k_0 = \frac{3\pi \beta k_\mathrm{B} T_\mathrm{e}}{c^2}\frac{D_\mathrm{cs}
}{D_\mathrm{s}}\frac{1}{\theta_\mathrm{c,proj}}\left(
1+e_\mathrm{proj}^2\right) {\times}
\left\{
\begin{tabular}{ll}
${\displaystyle \frac{e_\mathrm{int}}{e_{\rm proj}^2 } }$ & {\rm if prolate}
\\ ${\displaystyle \frac{e_{\rm proj}}{e_\mathrm{int}} }$ & {\rm if oblate}  \\
\end{tabular}
\right.
\eeq
where $D_\mathrm{cs}$ and $D_\mathrm{s}$ are the angular diameter distances from the deflecting cluster to the lensed background source and from the observer to the source, respectively.

The equation for the convergence can be coupled to those discussed in
\S~\ref{sec:elli} to determine if the cluster is oblate or prolate,
and to estimate one cosmological parameter, in particular $H_0$,
together with the shape parameters $i$ and $e_\mathrm{int}$.
Additional multiple image systems allow us to estimate further
cosmological parameters (see~\citetalias{def+al05}).

\section{Systematics}
\label{sec:syst}

Several systematic uncertainties that depend on the physical state of
the ICM could affect our determination of the intrinsic structure of a
galaxy cluster. Let us review some effects.

\subsection{Cooling flows and temperature gradients}

Departures from isothermality such as a non constant temperature
profile or temperature sub-structures mainly affect our analysis
through the SZE, since $\Delta T \propto \int n_\mathrm{e} T dl$.

ICM temperature profiles have been accurately measured thanks to new
generation X-ray satellites for several nearby relaxed clusters
~\citep{pif+al05,vik+al05a}, although, due to restricted field of view
or background levels, in most cases such measurements are only
possible out to one-third/half of the virial radius. Results are still
controversial, but some recent investigations seemed to find a similar
behaviour in the ICM temperature distribution, with a broad peak
followed by a decrease at larger radii~\citep{pif+al05,vik+al05b}.
Cooling processes also alter the temperature profiles with sharp drops
within cluster cores~\citep{all+al01}. Similar measurements for large
samples of more distant clusters are instead still lacking.

Since the SZE depends essentially on the pressure profile of the
cluster, the presence of a cooling flow can affect our estimate of the
cluster extent along the line of sight. Our assumption of
isothermality in the X-ray analysis generally leads to an overestimate
of the temperature and (to a lesser extent) the pressure in the cool
region. This in turn leads to an overestimate of the expected SZE
signal. The observed temperature decrement is thus misinterpreted,
leading to an underestimate of the LOS extent.

A refined view of cooling processes should account for changes in
density profiles as well as the temperature profile, with the effect
due to decrease of gas temperature moderated by the sudden increase of
the gas density. Previous studies addressed the effect of departures
from isothermality when using SZE along with X-ray observations. Given
our different approach, we interpret as an over (under)-estimation of
the LOS elongation what has been previously seen as an under
(over)-estimation of the Hubble constant. Based on theoretical models
for radiative cooling, it was found that, even after excluding $\sim
80\%$ of the cooling-flow region, an error of $\sim 10\%$ on the
elongation could affect analyses restricted to radii of $1.5\
r_\mathrm{c}$ \citep{ma+na00}.

In order to check how much the presence of a central cooling region
can affect our analysis, we re-analyzed three massive cooling flow
clusters in our sample (RX~J1347.5-1145, A~1835 and A~478) using the
most recent {\it Chandra} observations available. We performed
morphological and spectral analyses both using the whole cluster
extent and excluding the central cool region. Details on the data
reduction and on the morphological analysis are given in
\citetalias{def+al05}. Removing the central region leads to an
increase of both values of $\theta_\mathrm{c}$ ($\approx 60\%$) and
$\beta$ ($\gs 6\%$), when fitting a $\beta$-model to the cluster
surface brightness elliptical profile. Estimates of central amplitude
are strongly affected too. Average isothermal spectral temperatures
are larger by $\sim 20\%$ respect to when the values are averaged over
the whole cluster extent. The above effects leads to an
underestimation of the cluster elongation along the line of sight of
$\sim 15\%$, in agreement with previous
estimates~\citep{ina+al95,ma+na00}.

The effect of a central decrement is opposite to that of a negative
gradient at large radii in the temperature profile when determining
the cluster structure assuming isothermality
\citep{ina+al95,yos+al98}. If the gas temperature in the outer region
is lower than that in the inner region, an overestimation of the LOS
extent as large as $\sim 20\%$ is possible \citep{ina+al95}.

Actually, systematic errors due to departures from isothermality are
overwhelmed by large observational uncertainties in both the actual
temperature profile and the SZE measurement. Results turn out to be
quite insensitive to  details of the gas modelling and the isothermal
$\beta$-model can still provide a reasonably accurate model
\citep{bon+al05}. Effects due to either central drops or temperature
gradients at large radii could partially compensate each other and
could be combined in a systematic deviation as large as $15\%$, whose
sign depends on which effect prevails.

\subsection{Density clumping}

Small-scale density fluctuations on X-ray measurements arising from
accretion events and major mergers could introduce a further bias.
Clumping in the gas density distribution causes an enhancement of the
X-ray brightness by a factor $C_\mathrm{n}=\langle n_{\rm
e}^2\rangle/\langle n_{\rm e}\rangle^2$ with respect to a uniform
smooth atmosphere, while SZE measurements are not affected. This leads
to an underestimate of the elongation along the line of sight
 for $C_\mathrm{n}>1$ ($e_\mathrm{LOS} \propto C_\mathrm{n}^{-1} $) .

Currently, there is no observational evidence of significant clumping
in galaxy clusters \citep{ree+al02}, although numerical
hydro-simulations show that $C_\mathrm{n} \simeq 1.34$
\citep{mat+al99}. More detailed simulations and observations are
required to estimate the extent and the effect that clumpiness could
induce on our analysis.

%\subsection{Magnetic fields}

%A magnetic field in the cluster plasma might support a non thermal
%component in the X-ray emission. It could be effective in the core
%regions of the cluster but it is unlikely that it plays a major role
%out to the hydrostatic radius \citep{coo98}. As a general trend, a
%non-gravitational source of pressure in the central regions should
%induce an underestimation of the elongation.

\subsection{Overall departure}

In general, systematic effects do not act all in the same
direction and require a very detailed modelling in order to correct
for an amount that might be not very significant. Different
effects can combine to give a nearly null bias, as suggested from analysis of subsamples. In
particular, cooling flow clusters in our sample are equally split
between those elongated and compressed along the LOS, suggesting that
including a temperature profile or other effects would not drastically
change the results on their average structure.

Departures from an isothermal ICM in hydrostatic equilibrium should
not introduce serious errors. The reliability of isothermal models to
predict expected X-ray or SZE observations was tested against
high-resolution, hydrodynamic cluster simulations \citep{flo+al05}. As
long as one focuses on cluster regions that are less sensitive to
recent mergers, assuming isothermal gas allows accurate X-ray
estimations even if the gas has a strong temperature gradient. SZE
decrement maps can also be accounted for although not as successfully
as for the X-ray case \citep{flo+al05}.

\section{Discussion and Conclusions}
\label{sec:conc}

We have studied the intrinsic structure of clusters of galaxies by
combining X-ray and SZE observations and modelling them as ellipsoids
of revolution. We find that clusters of galaxies cannot be described
as a population of either purely oblate or purely prolate spheroids
with a random inclination of the symmetry axis respect to the LOS. A
mixed population is instead required to model the data, prolate-like
shapes being slightly preferred.

We have then compared our results with what is expected from a
population of randomly oriented galaxy clusters, having the observed
projected ellipticities of our sample. In contrast with the
intrinsic population expected under this hypothesis, we find a significant excess of clusters nearly aligned along the line of sight. Furthermore, high inclinations angles
are observed mainly for oblate clusters, while most elongated clusters turn out to be only prolate. If we believe the hypothesis of randomly
oriented polar axes to be correct, the behavior of the clusters in our
sample could be explained by a combination of two effects: a first one
caused by selection effects, which favours the detection of clusters
more elongated along the line of sight, and a second one due to the
fact that an oblate or prolate ellipsoidal model might not be correct for galaxy
clusters, and a more general triaxial morphology should instead be
used. Even though clusters in our sample were originally selected
according to their luminosity with a selection threshold well above
the detection limit, a selection bias on the basis of X-ray surface brightness can still persists for extremely elongated clusters. The second effect would instead lead to more dramatic conclusions.

Higher accuracy measurements and a better understanding of systematics
are needed to confirm our results. Although approximating galaxy
clusters as isothermal systems in hydrostatic equilibrium with
ellipsoidal matter distributions can be quite simplistic for some
clusters, it can give a first insight on their intrinsic structure.
Some numerical simulations also seem to support this view
\citep{flo+al05}. A natural development could be using more
sophisticated models than the isothermal-$\beta$ profile to describe
the ICM. Due to the infinite extent assumed in this model, the slope
of the surface brightness distribution at large radii could be too
shallow and might miss a progressive steepening with radius.
Unfortunately, data of sufficient quality are often missing too and
ICM can be traced with sufficient accuracy only up to few core radii.
Furthermore, the $\beta$-model provide a very good framework for the
gas density distribution of most of the observed clusters in our
sample and still provide a bench-mark for comparison in a statistical
sample. High-resolution spatially resolved spectral measurements of
temperature profile are restricted within the very inner regions for
most of the clusters and some extrapolation technique, with a degree
of unavoidable liberty, is therefore required \citep{sch+al04}.
Furthermore, from recent studies there appeared to be little
difference in the accuracy of the mass estimation from X-ray or SZE
measurements whether using isothermal or general, numerically or
observationally motivated, temperature profile methods
\citep{hal+al05}. While waiting for more precise measurements on a
cluster-by-cluster basis, using only a few numbers, i.e. a single
temperature, a constant axial ratio and a simple model for the ICM
distribution, is still a conservative approach. A crucial step to strengthen the statistical significance of our analysis would be the use of an unbiased and larger sample. The sample we have been considering in this paper is restricted to X-ray selected clusters for which a SZE analysis had been already reported. Preferential inclusion of high-S/N clusters could bias the sample towards clusters highly elongated towards the LOS.

The problem of determining the intrinsic shape of galaxy clusters
could be in principle completely solved for quite general
morphologies, but uncertainties on observed quantities make this task
still hard to obtain. An improved accuracy in the measurement of the SZE would also allow to look for trends in the intrinsic structure with mass or redshift as shown by numerical simulations \citep{ji+su02,ka+ev05,paz+al06}. Unfortunately, present data are not sufficiently precise to test such predicted correlations \citep{def+al05}. As we have discussed in~\citetalias{def+al05}, a proper triaxial structure is well suited to fit observations. Additional independent constraints from gravitational lensing
observations can break the degeneracy on the intrinsic shape of galaxy
clusters and discriminate between triaxial spheroids and ellipsoids of
revolution. In view of future progresses on the observational side, a
more accurate adaptation of our method, in which both the assumptions
of isothermality and of a density profile approximated by a
$\beta$-model are abandoned, will be presented in a forthcoming paper.
Our complete 3-D deprojection method will also be investigated with
numerical simulations. The comparison of the projected properties of 3-D simulated haloes with observational data could help to understand the intrinsic structure and to test the approximations and hypotheses used \citep{paz+al06}. With respect to similar previous analyses of groups based on the optical distribution of member galaxies, our method would have the advantage to be not affected by systematics due to finite sampling \citep{paz+al06}.

\acknowledgements
The authors thank the referee for the useful suggestions. This work has been partially supported by MIUR grant COFIN~2004020323 and by NASA grants NAS8-39073 and NAS8-00128. M.S. is supported by the Swiss National Science Foundation and by the Tomalla Foundation. This research has made use of NASA's Astrophysics Data system Bibliographic Services.

\appendix
\section{Two-dimensional projection of a three-dimensional ellipsoid}
\label{sec:appendiceA}

The expressions reported in Sections~\ref{sec:elli} and \ref{sec:lens}
can be obtained from the results in \citetalias{def+al05}. Let us
denote the observer coordinate system as $\left\{ x_{i,\rm obs}
\right\}$, $i=1,2,3$. The polar axis forms an angle $i$ with the line
of sight, $x_{3,\rm obs}$. We assume that the major axis of the
projected ellipses always lies along the $x_{1,\rm obs}$-axis, i.e.
the polar axis lies in the $x_{1,\rm obs}-x_{3,\rm obs}$ plane in the
prolate case and in the $x_{2,\rm obs}-x_{3,\rm obs}$ plane in the
oblate case. As an example, in the prolate case, the relation between
the two coordinate systems are
\begin{eqnarray}
x_{1,\rm int} & = &  x_{1,\rm obs} \cos i + x_{3,\rm obs}\sin i,
\\
x_{2,\rm int} & = &  x_{2,\rm obs},
\\ x_{3,\rm int} & = & x_{3,\rm obs} \cos i - x_{1,\rm obs} \sin i .
\end{eqnarray}
Following the notation in \citetalias{def+al05}, for a prolate
ellipsoid $v_1=v_2=e_\mathrm{int}$, $\theta_\mathrm{c3}=\theta_c$; for
an oblate ellipsoid $v_1=v_2=1/e_\mathrm{int}$,
$\theta_\mathrm{c3}=\theta_c/e_\mathrm{int}$. The observed core radius
is
\beq
\label{obl1}
\theta_{\rm proj} = \theta_c {\times}
\left\{
\begin{array}{ll}
{\displaystyle \frac{e_\mathrm{proj}}{e_\mathrm{int}} } & \mathrm{prolate\ case}
\\
1 & \mathrm{oblate\ case}  \\
\end{array}
\right.
\eeq
Substituting in the formulae reported in \citetalias{def+al05}, we
obtain the expressions used in Sections~\ref{sec:elli} and
\ref{sec:lens}.

%\bibliographystyle{apj}

%\bibliography{apj-jour,biblio_inclination}

\begin{thebibliography}{53}
\expandafter\ifx\csname natexlab\endcsname\relax\def\natexlab#1{#1}\fi

\bibitem[{{Alam} \& {Ryden}(2002)}]{al+ry02}
{Alam}, S.~M.~K., \& {Ryden}, B.~S. 2002, \apj, 570, 610

\bibitem[{{Allen} {et~al.}(2001){Allen}, {Schmidt}, \& {Fabian}}]{all+al01}
{Allen}, S.~W., {Schmidt}, R.~W., \& {Fabian}, A.~C. 2001, \mnras, 328, L37

\bibitem[{{Basilakos} {et~al.}(2000){Basilakos}, {Plionis}, \&
  {Maddox}}]{bas+al00}
{Basilakos}, S., {Plionis}, M., \& {Maddox}, S.~J. 2000, \mnras, 316, 779

\bibitem[{{Bertola} \& {Capaccioli}(1975)}]{be+ca75}
{Bertola}, F., \& {Capaccioli}, M. 1975, \apj, 200, 439

\bibitem[{{Binggeli}(1980)}]{bin80}
{Binggeli}, B. 1980, A\&A, 82, 289

\bibitem[{{Binney} \& {de Vaucouleurs}(1981)}]{bi+de81}
{Binney}, J., \& {de Vaucouleurs}, G. 1981, \mnras, 194, 679

\bibitem[{{Binney} \& {Merrifield}(1998)}]{bi+me98}
{Binney}, J., \& {Merrifield}, M. 1998, {Galactic astronomy} (Princeton:
  Princeton University Press)

\bibitem[{{Bonamente} {et~al.}(2005){Bonamente}, {Joy}, {LaRoque}, {Carlstrom},
  {Reese}, \& {Dawson}}]{bon+al05}
{Bonamente}, M., {Joy}, M., {LaRoque}, S.~J., {Carlstrom}, J.~E., {Reese},
  E.~D., \& {Dawson}, K.~S. 2005, astro-ph/0512349

\bibitem[{{Carter} \& {Metcalfe}(1980)}]{ca+me80}
{Carter}, D., \& {Metcalfe}, N. 1980, \mnras, 191, 325

\bibitem[{{Cooray}(1998)}]{coo98}
{Cooray}, A.~R. 1998, A\&A, 333, L71

\bibitem[{{Cooray}(2000)}]{coo00}
---. 2000, \mnras, 313, 783

\bibitem[{{D'Agostini}(2004)}]{dag04}
{D'Agostini}, G. 2004, physics/0403086

\bibitem[{{De Filippis} {et~al.}(2005){De Filippis}, {Sereno}, {Bautz}, \&
  {Longo}}]{def+al05}
{De Filippis}, E., {Sereno}, M., {Bautz}, M.~W., \& {Longo}, G. 2005, ApJ, in
  press; astro-ph/0502153

\bibitem[{{de Theije} {et~al.}(1995){de Theije}, {Katgert}, \& {van
  Kampen}}]{det+al95}
{de Theije}, P.~A.~M., {Katgert}, P., \& {van Kampen}, E. 1995, \mnras, 273, 30

\bibitem[{{Donahue} {et~al.}(2003){Donahue}, {Gaskin}, {Patel}, {Joy}, {Clowe},
  \& {Hughes}}]{don+al03}
{Donahue}, M., {Gaskin}, J.~A., {Patel}, S.~K., {Joy}, M., {Clowe}, D., \&
  {Hughes}, J.~P. 2003, \apj, 598, 190

\bibitem[{{Fabricant} {et~al.}(1984){Fabricant}, {Rybicki}, \&
  {Gorenstein}}]{fab+al84}
{Fabricant}, D., {Rybicki}, G., \& {Gorenstein}, P. 1984, \apj, 286, 186

\bibitem[{{Fasano} \& {Vio}(1991)}]{fa+vi91}
{Fasano}, G., \& {Vio}, R. 1991, \mnras, 249, 629

\bibitem[{{Fixsen} {et~al.}(1996){Fixsen}, {Cheng}, {Gales}, {Mather},
  {Shafer}, \& {Wright}}]{fix+al96}
{Fixsen}, D.~J., {Cheng}, E.~S., {Gales}, J.~M., {Mather}, J.~C., {Shafer},
  R.~A., \& {Wright}, E.~L. 1996, \apj, 473, 576

\bibitem[{{Flores} {et~al.}(2005){Flores}, {Allgood}, {Kravtsov}, {Primack},
  {Buote}, \& {Bullock}}]{flo+al05}
{Flores}, R.~A., {Allgood}, B., {Kravtsov}, A.~V., {Primack}, J.~R., {Buote},
  D.~A., \& {Bullock}, J.~S. 2005, ArXiv Astrophysics e-prints

\bibitem[{{Fox} \& {Pen}(2002)}]{fo+pe02}
{Fox}, D.~C., \& {Pen}, U. 2002, \apj, 574, 38

\bibitem[{{Hallman} {et~al.}(2005){Hallman}, {Motl}, {Burns}, \&
  {Norman}}]{hal+al05}
{Hallman}, E.~J., {Motl}, P.~M., {Burns}, J.~O., \& {Norman}, M.~L. 2005,
  astro-ph/0509460

\bibitem[{{Hubble}(1926)}]{hub26}
{Hubble}, E.~P. 1926, \apj, 64, 321

\bibitem[{{Illingworth}(1977)}]{ill77}
{Illingworth}, G. 1977, \apjl, 218, L43

\bibitem[{{Inagaki} {et~al.}(1995){Inagaki}, {Suginohara}, \&
  {Suto}}]{ina+al95}
{Inagaki}, Y., {Suginohara}, T., \& {Suto}, Y. 1995, \pasj, 47, 411

\bibitem[{{Jing} \& {Suto}(2002)}]{ji+su02}
{Jing}, Y.~P., \& {Suto}, Y. 2002, \apj, 574, 538

\bibitem[{{Kasun} \& {Evrard}(2005)}]{ka+ev05}
{Kasun}, S.~F., \& {Evrard}, A.~E. 2005, \apj, 629, 781

\bibitem[{{Kazantzidis} {et~al.}(2004){Kazantzidis}, {Kravtsov}, {Zentner},
  {Allgood}, {Nagai}, \& {Moore}}]{kaz+al04}
{Kazantzidis}, S., {Kravtsov}, A.~V., {Zentner}, A.~R., {Allgood}, B., {Nagai},
  D., \& {Moore}, B. 2004, \apjl, 611, L73

\bibitem[{{Majumdar} \& {Nath}(2000)}]{ma+na00}
{Majumdar}, S., \& {Nath}, B.~B. 2000, \apj, 542, 597

\bibitem[{{Mason} {et~al.}(2001){Mason}, {Myers}, \& {Readhead}}]{mas+al01}
{Mason}, B.~S., {Myers}, S.~T., \& {Readhead}, A.~C.~S. 2001, \apjl, 555, L11

\bibitem[{{Mathiesen} {et~al.}(1999){Mathiesen}, {Evrard}, \&
  {Mohr}}]{mat+al99}
{Mathiesen}, B., {Evrard}, A.~E., \& {Mohr}, J.~J. 1999, \apjl, 520, L21

\bibitem[{{Mohr} {et~al.}(1995){Mohr}, {Evrard}, {Fabricant}, \&
  {Geller}}]{moh+al95}
{Mohr}, J.~J., {Evrard}, A.~E., {Fabricant}, D.~G., \& {Geller}, M.~J. 1995,
  \apj, 447, 8

\bibitem[{{Noerdlinger}(1979)}]{noe79}
{Noerdlinger}, P.~D. 1979, \apj, 234, 802

\bibitem[{{Paz} {et~al.}(2006){Paz}, {Lambas}, {Padilla}, \&
  {Merch{\'a}n}}]{paz+al06}
{Paz}, D.~J., {Lambas}, D.~G., {Padilla}, N., \& {Merch{\'a}n}, M. 2006, \mnras
  ~in press, astro-ph/0509062

\bibitem[{{Piffaretti} {et~al.}(2005){Piffaretti}, {Jetzer}, {Kaastra}, \&
  {Tamura}}]{pif+al05}
{Piffaretti}, R., {Jetzer}, P., {Kaastra}, J.~S., \& {Tamura}, T. 2005, {\aa}p,
  433, 101

\bibitem[{{Plionis} {et~al.}(1991){Plionis}, {Barrow}, \& {Frenk}}]{pli+al91}
{Plionis}, M., {Barrow}, J.~D., \& {Frenk}, C.~S. 1991, \mnras, 249, 662

\bibitem[{{Plionis} {et~al.}(2004){Plionis}, {Basilakos}, \&
  {Tovmassian}}]{pli+al04}
{Plionis}, M., {Basilakos}, S., \& {Tovmassian}, H.~M. 2004, MNRAS, 352, 1323

\bibitem[{{Reblinsky}(2000)}]{reb00}
{Reblinsky}, K. 2000, A\&A, 364, 377

\bibitem[{{Reese} {et~al.}(2002){Reese}, {Carlstrom}, {Joy}, {Mohr}, {Grego},
  \& {Holzapfel}}]{ree+al02}
{Reese}, E.~D., {Carlstrom}, J.~E., {Joy}, M., {Mohr}, J.~J., {Grego}, L., \&
  {Holzapfel}, W.~L. 2002, \apj, 581, 53

\bibitem[{{Ryden}(1992)}]{ryd92}
{Ryden}, B. 1992, \apj, 396, 445

\bibitem[{{Ryden}(1996)}]{ryd96}
{Ryden}, B.~S. 1996, \apj, 461, 146

\bibitem[{{Schmidt} {et~al.}(2004){Schmidt}, {Allen}, \& {Fabian}}]{sch+al04}
{Schmidt}, R.~W., {Allen}, S.~W., \& {Fabian}, A.~C. 2004, \mnras, 352, 1413

\bibitem[{{Sereno} {et~al.}(2001){Sereno}, {Covone}, {Piedipalumbo}, \& {de
  Ritis}}]{ser+al01}
{Sereno}, M., {Covone}, G., {Piedipalumbo}, E., \& {de Ritis}, R. 2001, \mnras,
  327, 517

\bibitem[{{Tegmark} {et~al.}(2004){Tegmark}, {Strauss}, {Blanton}, {Abazajian},
  {Dodelson}, {Sandvik}, {Wang}, \& {Weinberg}}]{teg+al04}
{Tegmark}, M., {Strauss}, M.~A., {Blanton}, M.~R., {Abazajian}, K., {Dodelson},
  S., {Sandvik}, H., {Wang}, X., \& {Weinberg}, D.~H. e.~a. 2004, \prd, 69,
  103501

\bibitem[{{Thakur} \& {Chakraborty}(2001)}]{th+ch01}
{Thakur}, P., \& {Chakraborty}, D.~K. 2001, \mnras, 328, 330

\bibitem[{{Vikhlinin} {et~al.}(2005{\natexlab{a}}){Vikhlinin}, {Kravtsov},
  {Forman}, {Jones}, {Markevitch}, {Murray}, \& {Van Speybroeck}}]{vik+al05b}
{Vikhlinin}, A., {Kravtsov}, A., {Forman}, W., {Jones}, C., {Markevitch}, M.,
  {Murray}, S.~S., \& {Van Speybroeck}, L. 2005{\natexlab{a}}, ArXiv
  Astrophysics e-prints

\bibitem[{{Vikhlinin} {et~al.}(2005{\natexlab{b}}){Vikhlinin}, {Markevitch},
  {Murray}, {Jones}, {Forman}, \& {Van Speybroeck}}]{vik+al05a}
{Vikhlinin}, A., {Markevitch}, M., {Murray}, S.~S., {Jones}, C., {Forman}, W.,
  \& {Van Speybroeck}, L. 2005{\natexlab{b}}, \apj, 628, 655

\bibitem[{{Vio} {et~al.}(1994){Vio}, {Fasano}, {Lazzarin}, \&
  {Lessi}}]{vio+al94}
{Vio}, R., {Fasano}, G., {Lazzarin}, M., \& {Lessi}, O. 1994, A\&A, 289, 640

\bibitem[{{Wang} {et~al.}(2000){Wang}, {Caldwell}, {Ostriker}, \&
  {Steinhardt}}]{wan+al00}
{Wang}, L., {Caldwell}, R.~R., {Ostriker}, J.~P., \& {Steinhardt}, P.~J. 2000,
  \apj, 530, 17

\bibitem[{{Wang} \& {Fan}(2004)}]{wa+fa04}
{Wang}, Y.-G., \& {Fan}, Z.-H. 2004, \apj, 617, 847

\bibitem[{{West}(1994)}]{wes94}
{West}, M.~J. 1994, \mnras, 268, 79

\bibitem[{{Yoshikawa} {et~al.}(1998){Yoshikawa}, {Itoh}, \& {Suto}}]{yos+al98}
{Yoshikawa}, K., {Itoh}, M., \& {Suto}, Y. 1998, \pasj, 50, 203

\bibitem[{{Zaroubi} {et~al.}(2001){Zaroubi}, {Squires}, {de Gasperis},
  {Evrard}, {Hoffman}, \& {Silk}}]{zar+al01}
{Zaroubi}, S., {Squires}, G., {de Gasperis}, G., {Evrard}, A.~E., {Hoffman},
  Y., \& {Silk}, J. 2001, \apj, 561, 600

\bibitem[{{Zaroubi} {et~al.}(1998){Zaroubi}, {Squires}, {Hoffman}, \&
  {Silk}}]{zar+al98}
{Zaroubi}, S., {Squires}, G., {Hoffman}, Y., \& {Silk}, J. 1998, \apjl, 500,
  L87

\end{thebibliography}

\end{document}